\documentclass{ws-p10x7}

\def\slashchar#1{\setbox0=\hbox{$#1$}\dimen0=\wd0%
\setbox1=\hbox{/}\dimen1=\wd1%
\ifdim\dimen0>\dimen1%
\rlap{\hbox to 
\dimen0{\hfil/\hfil}}#1\else                                        
\rlap{\hbox to \dimen1{\hfil$#1$\hfil}}/\fi} 
\def\Journal#1#2#3#4{{#1} {\bf #2}, #3 (#4)}

\def\NPB{{\em Nucl. Phys.} B}

\def\PLB{{\em Phys. Lett.}  B}
\def\PRL{\em Phys. Rev. Lett.}
\def\PRD{{\em Phys. Rev.} D}
\def\ZPC{{\em Z. Phys.} C}

\newcommand{\as}{\alpha_{s}}  
\def\vepp{\varepsilon'}
\def\vep{\varepsilon}

\def\be{\begin{equation}}
\def\ee{\end{equation}}
\def\ba{\begin{eqnarray}}
\def\ea{\end{eqnarray}}

\def\cO{{\cal O}}

\def\cA{{\cal A}}
\begin{document}

\title{The Standard Model prediction for $\vepp / \vep$ }

\author{E. Pallante}

\address{Facultat de F\'{\i}sica, Universitat de Barcelona, Diagonal
647, E-08028 Barcelona, Spain, 
E-mail: pallante@ecm.ub.es}

\author{A. Pich and I. Scimemi}

\address{ Departament de F\'{\i}sica Te\`orica, IFIC,  
  Apt. Correus 2085, E--46071, Val\`encia,  Spain,  
E-mail: Antonio.Pich@uv.es, scimemi@hal.ific.uv.es}  

\twocolumn[\maketitle\abstract{
We briefly review the most important ingredients
 of a new Standard Model analysis of  
$\vepp / \vep$ which takes into account  
the strong enhancement induced by final state interactions.
}]

\section{Introduction}
\label{INTRO}

The study of non--leptonic $K\to \pi\pi$ decays is of great importance in 
the understanding of CP violation mechanisms within the Standard Model and 
beyond. In particular, a crucial quantity is the parameter 
$\vepp / \vep$ which measures the magnitude of the direct CP violation
in the Kaon system. The experimental situation has been greatly improved
recently, after the measurement by NA48 at CERN and 
KTeV at Fermilab. The new quoted experimental world average\cite{EXP}
is ${ \rm Re} \left(\vepp/\vep\right) =(19.3 \pm 2.4) \cdot 10^{-4}$,
providing a clear evidence of the existence of direct CP violation with 
a non--zero value of $\vepp / \vep$.

The theoretical prediction of $\vepp / \vep$ still suffers from many 
uncertainties which mainly affect the determinaton of the long--distance 
contributions to $K\to \pi\pi$ matrix elements and the matching with the 
short--distance part.
Recently, it has been observed\cite{SHORT,FUTURE_PP} that the
soft final state interactions (FSI) of the two pions 
play an important role 
in the determination of $\vepp / \vep$. From the measured $\pi$--$\pi$ phase 
shifts one can easily infer that FSI generate a strong enhancement of the 
predicted value of $\vepp / \vep$ by roughly a factor of 
two\cite{SHORT,FUTURE_PP}, providing a good  agreement with the 
experimental value.

Here, we discuss a few basic aspects of a new Standard Model evaluation
of $\vepp / \vep$ which has been proposed in 
Refs.\cite{SHORT,FUTURE_PP,FUTURE_PPS} and includes FSI
effects. In addition to the large infrared logarithms generated 
by FSI there are the well known large ultraviolet logarithms that 
govern the short--distance evolution of the Wilson coefficients.
Both these logarithms need to be resummed and included in the evaluation of 
$\vepp / \vep$. 
The large--$N_C$ expansion\cite{tHO:74,WI:79} provides a convenient
framework, with a well defined power counting, to properly include 
all these corrections.

In Sec.~\ref{KAON} we review the calculation of long--distance 
$K\to\pi\pi$ matrix elements with the inclusion of FSI effects, while 
Sec.~\ref{sec:EPS} is devoted to the evaluation of $\vepp / \vep$. 

\section{$K\to\pi\pi$ matrix elements}
\label{KAON}

The long--distance realization of matrix elements among
light pseudoscalar mesons 
can be obtained within Chiral Perturbation Theory (ChPT),
as an expansion in powers of momenta   
and light quark masses\cite{FUTURE_PP}.
The $K\to\pi\pi$ amplitudes with $I=0,2$ generated by the lowest--order 
ChPT lagrangian are
\ba
&& A_0\, =\, 
-{G_F\over \sqrt{2}} V_{ud}V^\ast_{us}\,\sqrt{2} f\,
 \nonumber\\
&&
\left\{\left(g_8+{1\over 9}\, g_{27}\right) (M_K-M_\pi^2) 
 -{2\over 3} f^2 e^2 g_{EM}\right\}  ,
\nonumber\\
&& A_2\, =\,
  -{G_F\over \sqrt{2}} V_{ud}V^\ast_{us}\, {2\over 9} f
\, \left\{5\, g_{27}\, (M_K-M_\pi^2)\, \right .\nonumber\\
&&\left .- 3 f^2 e^2 g_{EM}\right\} \, ,
\label{TREE}
\ea
where $g_8$, $g_{27}$ and $g_{EM}$ are the chiral couplings and 
 the isospin decomposition of Ref.\cite{FUTURE_PP} has been used.
 
FSI start at next--to--leading order in the chiral expansion.
To resum those effects the Omn\`es approach for $K\to\pi\pi$ decays has been 
proposed in Ref.\cite{SHORT} and discussed in detail in 
Ref.\cite{FUTURE_PP}. 
For CP--conserving amplitudes, where the $e^2 g_{EM}$ corrections can be safely 
neglected, the most general Omn\`es solution for 
the on--shell amplitude can be written as follows 
\ba\label{eq:OMNES_WA} 
&&\cA_I \, =\,   
\left(M_K^2-M_\pi^2\right) \; \Omega_I(M_K^2,s_0) \; a_I(s_0)\\
&&\, = \,
\left(M_K^2-M_\pi^2\right) \; \Re_I(M_K^2,s_0) \; a_I(s_0) 
\; e^{i\delta^I_0(M_K^2)}\, . 
\nonumber\ea 
The Omn\`es factor $\Omega_I(M_K^2,s_0)$ provides an evolution 
of the amplitude from low energy values (the subtraction point $s_0$), 
where the ChPT momentum expansion can be trusted, to higher energy values, 
through the exponentiation of the infrared effects due to FSI.
It can be split into the dispersive 
contribution $\Re_I(M_K^2,s_0) $ and the phase shift 
exponential\footnote{For the electroweak penguin
operator $Q_8$, the lowest order chiral contribution is a constant 
proportional to $e^2 g_{EM}$, where the explicit $SU(3)$ breaking 
is induced by the quark charge matrix, instead of the term $M_K^2-M_\pi^2$.}.
Taking a low subtraction point $s_0 =0$, we have shown\cite{FUTURE_PP}
that one can just multiply the tree--level 
formulae (\ref{TREE}) with the experimentally determined Omn\`es 
exponentials \cite{FUTURE_PP}.
The two dispersive correction factors thus obtained \cite{FUTURE_PP} are 
$\Re_0(M_K^2,0)\,=\, 1.55 \pm 0.10$  and 
$\Re_2(M_K^2,0) \,=\, 0.92 \pm 0.03$.

The complete derivation of the Omn\`es solution for  $K\to\pi\pi$ amplitudes 
 makes use of Time--Reversal 
invariance, so that it can be strictly applied only to CP--conserving 
amplitudes.
However, working at the first order in the Fermi coupling, the CP--odd phase 
is fully contained in the ratio of CKM matrix elements 
$\tau = V_{td}\,V^*_{ts} / V_{ud}\,V^*_{us} $ which multiplies 
the short--distance Wilson coefficients. 
Thus, decomposing the isospin amplitude 
as $\cA_I = \cA_I^{CP} + \tau \,\cA_I^{\slashchar{CP}} $, the Omn\`es 
solution can be derived for the two amplitudes $ \cA_I^{CP} $ and
 $\cA_I^{\slashchar{CP}}$ which respect Time--Reversal invariance.
In a more standard notation,
$\mbox{Re}A_I \approx A_I^{CP}$ and
$\mbox{Im}A_I = \mbox{Im}(\tau )\, A_I^{\slashchar{CP}}$,
where the absorptive phases have been already factored out
through $\cA_I = A_I\, e^{i\delta_0^I}$.
\section{The parameter $\vepp / \vep$}
\label{sec:EPS}
The direct CP violation parameter $\vepp /\vep$ can be written in terms 
of the definite isospin $K\to\pi\pi$ amplitudes as follows
\begin{equation}
{\varepsilon^\prime\over\varepsilon} =
\; e^{i\Phi}\; {\omega\over \sqrt{2}\vert\epsilon\vert}\left [ 
{\mbox{Im}A_2\over\mbox{Re} A_2} - {\mbox{Im}A_0\over \mbox{Re} A_0} \right ]
\, , 
\label{EPS}
\end{equation}
where
the phase $\Phi = \Phi_{\varepsilon^\prime} -  \Phi_\varepsilon \simeq 0$
and $\omega = \mbox{Re}A_2/ \mbox{Re} A_0$.
Since the hadronic matrix elements are quite uncertain theoretically,
the CP--conserving amplitudes $\mbox{Re}A_I$, and thus
the factor  $\omega$, are usually set to their experimentally
determined values; this automatically includes the FSI effect.
All the rest in the numerator has been {\em theoretically} predicted mostly via
short--distance calculations, which therefore do not include FSI corrections.
This produces a mismatch which can be easily corrected by introducing 
in the numerator the appropriate dispersive factors $\Re_I$ for FSI effects.
The evaluation of $\vepp /\vep$ proposed in Ref.\cite{FUTURE_PPS} 
proceeds through the following steps:

\noindent $\bullet$ All short--distance Wilson coefficients are evolved 
at next-to-leading logarithmic order\cite{buras1,ciuc1} down to the charm 
quark mass scale $\mu = m_c$.  
All gluonic corrections of $\cO(\as^n t^n)$ and $\cO(\as^{n+1} t^n)$ 
are already known. Moreover, the full $m_t/M_W$ dependence (at lowest 
order in $\as$) has been taken into account. 
This provides the resummation of the large ultraviolet logarithms 
$t\equiv\ln{(M/m)}$, where $M$ and $m$ refer to any scales 
appearing in the evolution from $M_W$ down to $m_c$. 

\noindent $\bullet$ At the scale $\mu \sim 1$ GeV the $1/N_C$ expansion 
can be safely implemented. At this scale the logarithms which govern the 
evolution of the Wilson coefficients remain small $\sim\ln{(m_c/\mu )}$ so 
that the $1/N_C$ expansion has a clear meaning within the usual perturbative 
expansion in powers of $\alpha_s$. 
In the large-$N_C$ limit both the Wilson coefficients $C_i(\mu )$ and the 
long--distance matrix elements $\langle Q_i(\mu )\rangle_I$ can be computed 
and the matching at the scale $\mu \leq m_c$ can be done {\em exactly}.

\noindent $\bullet$ The Omn\`es procedure can be applied to the individual 
matrix elements $\langle Q_i\rangle_I$. Since the FSI effect is 
next--to--leading in the $1/N_C$ expansion one can include it via 
the realization $\langle Q_i(\mu )\rangle_I \sim 
\langle Q_i(\mu )\rangle_I^{N_C\to\infty} \times \Re_I$, while avoiding
any double counting.

The large-$N_C$ realization of the matrix elements
 $\langle Q_i(\mu )\rangle_I$, with $i\neq 6,8$ is always a product of 
the matrix elements of colour--singlet vector or axial currents. Each of them 
being an observable, the corresponding matrix element is renormalization 
scale and scheme independent. The same is true for the corresponding 
Wilson coefficients in the large--$N_C$ limit, so that the matching is exact.
The large--$N_C$ realization of $\langle Q_{6(8)}(\mu )\rangle_I$ 
scales like the inverse of the squared fermion mass, being  
 the product of colour--singlet scalar and pseudoscalar currents.
Conversely,  the Wilson coefficients of the operators 
$Q_6$ and $Q_8$ scale proportionally to the square of 
a quark mass in the large--$N_c$ limit, so that again the matching is exact.

The connection between the tree level ChPT amplitudes (\ref{TREE})
and the large--$N_C$ realization of the operators $Q_i$ can be clarified 
as follows.
At the lowest non trivial order in the chiral expansion, the large--$N_C$ 
realization of an operator $Q_i$ gives the contribution of $Q_i$ itself
to the chiral couplings $g_8$, or $g_{27}$ or $g_{EM}$ (according to its 
transformation properties) in the large--$N_C$ limit.
In this sense, the Omn\`es solution  formulated in Sec.~\ref{KAON} can 
be applied directly to the large--$N_C$ matrix elements of the operators $Q_i$
 with the dispersive factors $\Re_I(M_K^2,0)$ already estimated.
 
A preliminary Standard Model analysis of $\vepp / \vep$ gives
$\vepp / \vep = (17\pm 6)\cdot 10^{-4}$, where the error is dominated 
by the $1/N_C$ approximation. Further refinement and details of the analysis
will be given elsewhere\cite{FUTURE_PPS}. 


\end{document}